\documentclass[12pt]{article}
\usepackage{latexsym,graphicx}
\usepackage{subfigure}
\usepackage{amssymb}
\usepackage{amsmath}
\usepackage{amscd}
\usepackage{amsthm}
\usepackage{float}
\usepackage[left=2cm,top=2.5cm,right=2.5cm,bottom=1.5cm]{geometry}
\theoremstyle{definition}
\newtheorem{definition}{Definition}    
\newtheorem{theorem}{Theorem}[section]       
\begin{document}
\begin{center}
\large{\bf{Topological Origin of Holographic Principle: Application to wormholes}} \\
\vspace{5mm}
\normalsize{Nasr Ahmed$^a$ and H. Rafat$^b$,$^c$}\\
\vspace{5mm}
\small{\footnotesize$^{a}$ Astronomy Department, National Research Institute of Astronomy and Geophysics, Helwan, Cairo, Egypt\footnote{nasr.ahmed@nriag.sci.eg}} \\
\small{\footnotesize$^{b}$Department of Mathematics, Faculty of Science, Tanta University, Tanta, Egypt.\\$^c$Mathematics Department, Faculty of Science, Taibah University, Saudi Arabia} \\
\vspace{2mm}
\end{center}  
\begin{abstract}
In this paper, we suggest a mathematical representation to the holographic principle through the theory topological retracts. 
We found that the topological retraction is the mathematical analogs of the hologram idea in modern quantum gravity and it can be used to explore the geometry of the hologram boundary. An example has been given on the five dimensional (5D) wormhole space-time $W$ which we found it can retract to lower dimensional circles $S_i \subset W$. In terms of the holographic principle, the description of this volume of space-time $W$ is encoded on the lower-dimensional circle which is the region boundary.
\end{abstract}

\section{Introduction}

 In HP, information contained in a volume of space can be represented as a hologram and the 3 dimensional world is an image of data stored on a 2 dimensional projection \cite{tof,susskind}. This idea reduces the number of dimensions considered in a physical theory from $n$ to $n-1$ and, consequently, reduces the complexity of the physical theory. The holographic principle (HP) has become an extraordinary tool in theoretical physics. Most notably in the form of the Anti-deSitter/Conformal Field Theory (AdS/CFT) correspondence, in which quantum gravity in $n$ dimensional Anti-de Sitter space (AdS) is equivalent to a Conformal Field Theory (CFT) in $n-1$ dimensions defined on the boundary of the $n$ dimensional Anti-de Sitter space \cite{phd}.\\ In 1973, Bekenstein proposed that the entropy of a static black hole should be proportional to the event horizon area and not the volume \cite{beken}, $S_{bh} \propto \frac{A_{bh}}{2}.\frac{\ln 2}{4\pi}$. Soon after, the exact relation was determined by Hawking \cite{hawkk},
\begin{equation}
S_{BH} =\frac{kA}{4l_p^2}.
\end{equation}
Where $l_p=\sqrt{Gh/c^3}$ is the Planck length and $k$ is Boltzmann's constant,. In 1981, Bekenstein found an upper limit on the entropy $S$ (or information $I$) that can be contained in a physical system or object with given size and total energy \cite{bound},
\begin{equation}
S \leq \frac{2\pi k RE}{hc}.
\end{equation}
Where $R$ is the radius of a sphere that can enclose a given system and $E$ is its total energy including any rest masses. The fact that the limit on the entropy is set by the area and not
by the volume was surprising and led to the expectation that gravitational physics reduces the number of physical
degrees of freedom. It was then conjectured that quantum gravity should be described by a topological field theory, in the
sense that all its degrees of freedom live on the boundary \cite{haro}. \par

\section{Motivation}

In spite of the increasing importance of HP in theoretical physics, there is no solid mathematical base for this principle. The absence of a mathematical explanation for HP was the main motivation for this paper. Mathematically, the main idea behind HP is projecting a space of $n$ dimension into a subspace of $n-1$ dimensions and we have found that the topological retraction is a mathematical version of this physical principle. Although many topics in topology have found applications in theoretical physics, the retraction theory has not been used in physics before. Hence, the main purposes of this paper can be summarized in two points: 1- showing that HP has a strong base in pure mathematics. 2- providing a physical application to the retraction theory in algebraic topology.\par

Algebraic topology was considered a revolution in pure mathematics and theoretical physics in the second half of twentieth century. The concepts and methods of topology have become an indispensable part of theoretical physics and led to a deeper understanding of many crucial aspects in gravity, cosmology and particle physics \cite{geo}. In the 1890s, Poincar´e studied celestial mechanics by using topological results to prove the existence of periodic orbits \cite{vanessa}. In condensed matter, topology explains the precise quantization of the Hall effect \cite{hall}. In general relativity, topology enters through the assumption that space-time exists as a manifold. The topology change in general relativity has been discussed in \cite{topo2}. Some applications of differential topology in general relativity has been mentioned in \cite{topo3}. In quantum field theory (QFT), the effects of Topology occur due to the nontrivial global structure of the configuration space of the field theory. in the late 1970s, it was shown that QFT techniques can be used to obtain topological invariants of manifolds \cite{Witt,schwarz}. A definition of topological quantum field theory was given in \cite{ati}.

\section{Retraction and HP} \label{secsec}
\subsection{Retraction}

The theory of retraction is an interesting topic in Euclidean and non-Euclidean spaces. It has been investigated from different points of view in many branches of topology and differential geometry \cite{retra}.\\ A subspace $A$ of a topological space $X$ is called a retract of $X$, if there exists a continuous map $r : X \rightarrow A $ such that $X$ is open and $r(a) = a$ (identity map), $\forall a \in A$. \\Because the continuous map $r$ is an identity map from $X$ into $A \subset X$, it preserves the position of all points in $A$. Most of the studies on retractions, if not all, are pure mathematical studies. This paper shows a possible application to this topic in mathematical physics. Retractions of Stein spaces has been studied in \cite{stien}. Retraction of the Schwarzchild  metric has been discussed in \cite{hisham,nash}. A good example has been given by Kinsey \cite{kinsey} where we have a retraction $r:S^2 \rightarrow \left\{(1,0,0)\right\} \in S^2$ taking the sphere to a point defined by $r(x,y,z)=(1,0,0)$. When the retraction preserves the essential shape of the space, it is called deformation retract. For example, both the cylinder and the mobius band deformation retract to a circle and they are said to have the same homotopy type \cite{kinsey}.

\subsection{A connection to HP}

After we have shown the similarity between the retraction theory and HP where 'the projection onto a subspace' is the central idea in both of them, we need to set a clear statement describes this relation. On the math side, we have a pure topological idea of continuously shrinking a space $X$ into a subspace $A \in X$ which preserves the position of all points in that subspace \cite{top1}. On the physics side, we have the idea of a hologram where the description of a volume of space is encoded on a lower-dimensional subspace (boundary) to the region. In other words, retraction is the topological representation (a topological version) of HP, and HP is the physical representation (a physical version) of the topological retraction. This relation can be summarized in the following topological restatement to HP where the words between brackets represent the corresponding physical description:  

\theoremstyle{definition}
\begin{definition}{Topological restatement to HP:}
The holographic principle can be represented mathematically by topological retraction of $n$-dimensional space (hologram) to $n-1$ subspace (a boundary) that preserves the position of all points (preserving all information describing the $n$-dimensional space).
\end{definition}
A detailed example has been given in the next section where the retraction of a 5D Ricci-flat space has been performed and the hologram boundary geometry has been identified. 

\section{5D Wormhole metric}

The retraction method we are going to use in this section is applicable only for Ricci-flat geometries such as $n$D Schwarzchild space-time, Mcvittie space-time and Kerr space-time. A more general method will be required to perform the retractions of non Ricci-flat metrics. The $5$D Ricci-flat wormhole space-time $W$ is given by \cite{basic}:

\begin{equation}
ds^{2}=\frac{r^2-a^2}{r^2+a^2}(-dt^2+dz^2)+dr^2+(r^2+a^2)(d\theta^2+\sin^2\theta \,d\phi^2)-\frac{4ar}{r^2+a^2} dt\, dz. \label{mett}
\end{equation}

Let's briefly describe the geometry of wormholes. They are hypothetical bridges between two regions of the same universe (or different universes) connected by a throat. The throat of the wormholes is defined as a two dimensional hypersurface of minimal area. A wormhole is often referred to as Einstein-Rosen bridge because a "bridge" connecting two sheets was the result obtained by Einstein and Rosen in 1935 \cite{einst}. A wormhole solution is typically defined as a solution that is asymptotic to two distinct flat spacetimes.
Many interesting studies of Lorentzian and Euclidean wormholes have been produced in the past decade \cite{wormhole}. Lorentzian wormholes represent tiny quantum fluctuations of space and can join spaces with different topology \cite{wormtop}. Originally, wormholes were three-dimensional space. Four-dimensional and higher-dimensional wormholes were introduced by Hawking and Coleman \cite{h1,h2}. \par
 As indicated in \cite{basic}, the off-diagonal term in (\ref{mett}) can be removed through complex coordinate transformations. There are two asymptotic regions, corresponding to $r\rightarrow \pm \infty$, where the metric has the following form
\begin{equation}
ds^{2}=-dt^{2}+dz^{2}+dr^{2}+r^2 (d\theta^2+\sin^2\theta \,d\phi^2)
\end{equation}
This describes $(Minkowskian)_5$ if $z$ is a real line, or $(Minkowskian)_4 \times S^1$ if $z$ is a circle.
The general 5D flat metric has the form 
\begin{equation}
ds^{2}= -dx_{o}^{2}+ \sum_{i=1}^4 dx_{i}^{2}   \label{flatt}
\end{equation}
Comparing the flat metric (\ref{flatt}) with the Ricci-flat metric (\ref{mett}) and making use of the basic metric definition $ds^2 = g_{ij} dx^i dx^j$, the following coordinate relations can be obtained

\begin{eqnarray} \label{xi}
x_o&=&\pm \sqrt{\frac{r^2-a^2}{r^2+a^2} t^2+C_o}, \  \  \  x_1=\pm \sqrt{\frac{r^2-a^2}{r^2+a^2} z^2+C_1}, \  \  \  x_2=\pm \sqrt{r^2+C_2},\\  \nonumber
x_3&=&\pm \sqrt{(r^2+a^2)\theta^2+C_3}, \  \  \  x_4=\pm \sqrt{(r^2+a^2) \sin^2\theta \, \phi^2+C_4}.
\end{eqnarray}
Where $C_{o}$, $C_{1}$,$C_{2}$,$C_{3}$ and $C_{4}$ are constants of integration. These relations will play
a key role in the study of the geodesic retractions of the $5$D wormhole space-time $W$ (\ref{mett}). Now we need to find a geodesic which is a subset of $W$ and this can be done through the geodesic equation or the Euler-Lagrange equations. In general relativity, the geodesic equation is equivalent to the Euler-Lagrange equations 
\begin{equation}
\frac{d}{d\lambda}\left(\frac{\partial L}{\partial \dot{x}^{\alpha}}\right)-\frac{\partial L}{\partial x^{\alpha}} =0, ~~~~i=1,2,3,4
\end{equation}
associated to the Lagrangian 
\begin{equation}
L(x^{\mu},\dot{x}^{\mu})=\frac{1}{2}g_{\mu \nu}\dot{x}^{\mu}\dot{x}^{\nu}
\end{equation}
To find a geodesic which is a subset of $W$, we start from the lagrangian
\begin{equation}
L=-\frac{r^2-a^2}{2(r^2+a^2)}\dot{t}^2+\frac{r^2-a^2}{2(r^2+a^2)}\dot{z}^2+\frac{1}{2}\dot{r}^2 
\frac{1}{2}(r^2+a^2)(\dot{\theta}^2+\sin^2\theta \, \dot{\phi}^2)-\frac{2ar}{r^2+a^2} \dot{t} \dot{z} \label{Lagr}
\end{equation}
No explicit dependence on either $t$, $\phi$, or $z$ and thus $\frac{\partial L}{\partial \dot{t}}$, $\frac{\partial L}{\partial \dot{\phi}}$ and $\frac{\partial L}{\partial \dot{z}}$ are  constants of motion, i.e.
\begin{equation} \label{1}
-\frac{r^2-a^2}{r^2+a^2}\dot{t}=K. 
\end{equation}
\begin{equation}\label{2}
(r^2+a^2) \sin^2\theta \, \dot{\phi}=h.
\end{equation}
The $r$-component gives
\begin{equation}\label{3}
\frac{d}{d\lambda}(\dot{r})-\left[-\frac{2a^2r}{(r^2+a^2)^2}\dot{t}^2+\frac{2a^2r}{(r^2+a^2)^2}\dot{z}^2+r(\dot{\theta}^2+\sin^2\theta \, \dot{\phi}^2)+2a\frac{r^2-a^2}{(r^2+a^2)^2}\dot{t}\dot{z}\right]=0.
\end{equation}
The $\theta$-component gives
\begin{equation}\label{4}
\frac{d}{d\lambda}\dot{\theta}(r^2+a^2)-\left[\frac{1}{2}(r^2+a^2)\sin 2\theta \, \dot{\phi}^2\right]=0.
\end{equation}
The $\phi$-component gives
\begin{equation}\label{6}
\frac{d}{d\lambda}\left[(r^2+a^2)\dot{\phi}\sin^2\theta\right]=0.
\end{equation}
The $t$-component gives
\begin{equation}\label{7}
\frac{r^2-a^2}{r^2+a^2}\dot{t}+\frac{2ar}{r^2+a^2}\dot{z}=A.
\end{equation}
The $z$-component gives
\begin{equation}\label{8}
\frac{r^2-a^2}{r^2+a^2}\dot{z}-\frac{2ar}{r^2+a^2}\dot{t}=B.
\end{equation}
Where $A$ and $B$ are constants. Equations of motion for $\theta$ and $\phi$ admit the following solution:
\begin{equation}\label{5}
\theta=\frac{\pi}{2} , \,\,\, \dot{\phi}=\frac{D}{r^2+a^2}.
\end{equation}
Where $D$ is the integration constant describing the orbital angular momentum of the geodesic particle. Now we are going to make use of these equations to study the retraction of $W$ and see what types of geodesic retractions we obtain.

\section{Retraction}
From (\ref{1}), if $K=0$ this implies that $t=C$ where $C$ is a constant, or $r=a$. For $C=0$, the coordinates (\ref{xi}) become
\begin{eqnarray} \label{xi1}
x_o&=&\pm \sqrt{C_o}, \  \  \  x_1=\pm \sqrt{\frac{r^2-a^2}{r^2+a^2} z^2+C_1}, \  \  \  x_2=\pm \sqrt{r^2+C_2},\\  \nonumber
x_3&=&\pm \sqrt{(r^2+a^2)\theta^2+C_3}, \  \  \  x_4=\pm \sqrt{(r^2+a^2) \sin^2\theta \, \phi^2+C_4}.
\end{eqnarray}
Since $ds^2=x_{1}^{2}+x_{2}^{2}+x_{3}^{2}-x_{o}^{2}>0$ which is a circle $S_1 \subset W$. This geodesic is a retraction in the 5D space-time $W$ represented by the metric (\ref{mett}). For $r=a$, the coordinates (\ref{xi}) become
\begin{eqnarray} \label{xi2}
x_o&=&\pm \sqrt{C_o}, \  \  \  x_1=\pm \sqrt{C_1}, \  \  \  x_2=\pm \sqrt{r^2+C_2},\\  \nonumber x_3&=&\pm \sqrt{2a^2\theta^2+C_3}, \  \  \  x_4=\pm \sqrt{2a^2 \sin^2\theta \, \phi^2+C_4}.
\end{eqnarray}
Since $ds^2=x_{1}^{2}+x_{2}^{2}+x_{3}^{2}-x_{o}^{2}>0$ which is a circle $S_2 \subset W$. This geodesic is a retraction in the 5D space-time $W$ represented by the metric (\ref{mett}).\par
From (\ref{2}), if $h=0$ this implies that $\phi=H$ where $H$ is a constant, or $\theta=0$. For $H=0$, the coordinates (\ref{xi}) become
\begin{eqnarray} \label{xi3}
x_o&=&\pm \sqrt{\frac{r^2-a^2}{r^2+a^2} t^2+C_o}, \  \  \  x_1=\pm \sqrt{\frac{r^2-a^2}{r^2+a^2} z^2+C_1}, \  \  \  x_2=\pm \sqrt{r^2+C_2},\\  \nonumber
x_3&=&\pm \sqrt{(r^2+a^2)\theta^2+C_3}, \  \  \  x_4=\pm \sqrt{C_4}.
\end{eqnarray}
Since $ds^2=x_{1}^{2}+x_{2}^{2}+x_{3}^{2}-x_{o}^{2}>0$ which is a circle $S_3 \subset W$. This geodesic is a retraction in the 5D space-time $W$ represented by the metric (\ref{mett}). For $\theta=0$, we get
\begin{eqnarray} \label{xi4}
x_o&=&\pm \sqrt{\frac{r^2-a^2}{r^2+a^2} t^2+C_o}, \  \  \  x_1=\pm \sqrt{\frac{r^2-a^2}{r^2+a^2} z^2+C_1}, \  \  \  x_2=\pm \sqrt{r^2+C_2},\\  \nonumber
x_3&=&\pm \sqrt{C_3}, \  \  \  x_4=\pm \sqrt{C_4}.
\end{eqnarray}
Since $ds^2=x_{1}^{2}+x_{2}^{2}+x_{3}^{2}-x_{o}^{2}>0$ which is a circle $S_4 \subset W$. This geodesic is a retraction in the 5D space-time $W$ represented by the metric (\ref{mett}). So, the retraction of $W$ can now be defined as $R : W \Rightarrow S_i ,\,i=1,2,3,4$, and the following theorem has been proved
\begin{theorem}
Some types of the geodesic retractions of the 5D wormhole space-time are circles in the 5D wormhole space-time. 
\end{theorem}
\begin{figure}[H]
  \centering
  \subfigure[]{\label{energy0}\includegraphics[width=0.23\textwidth]{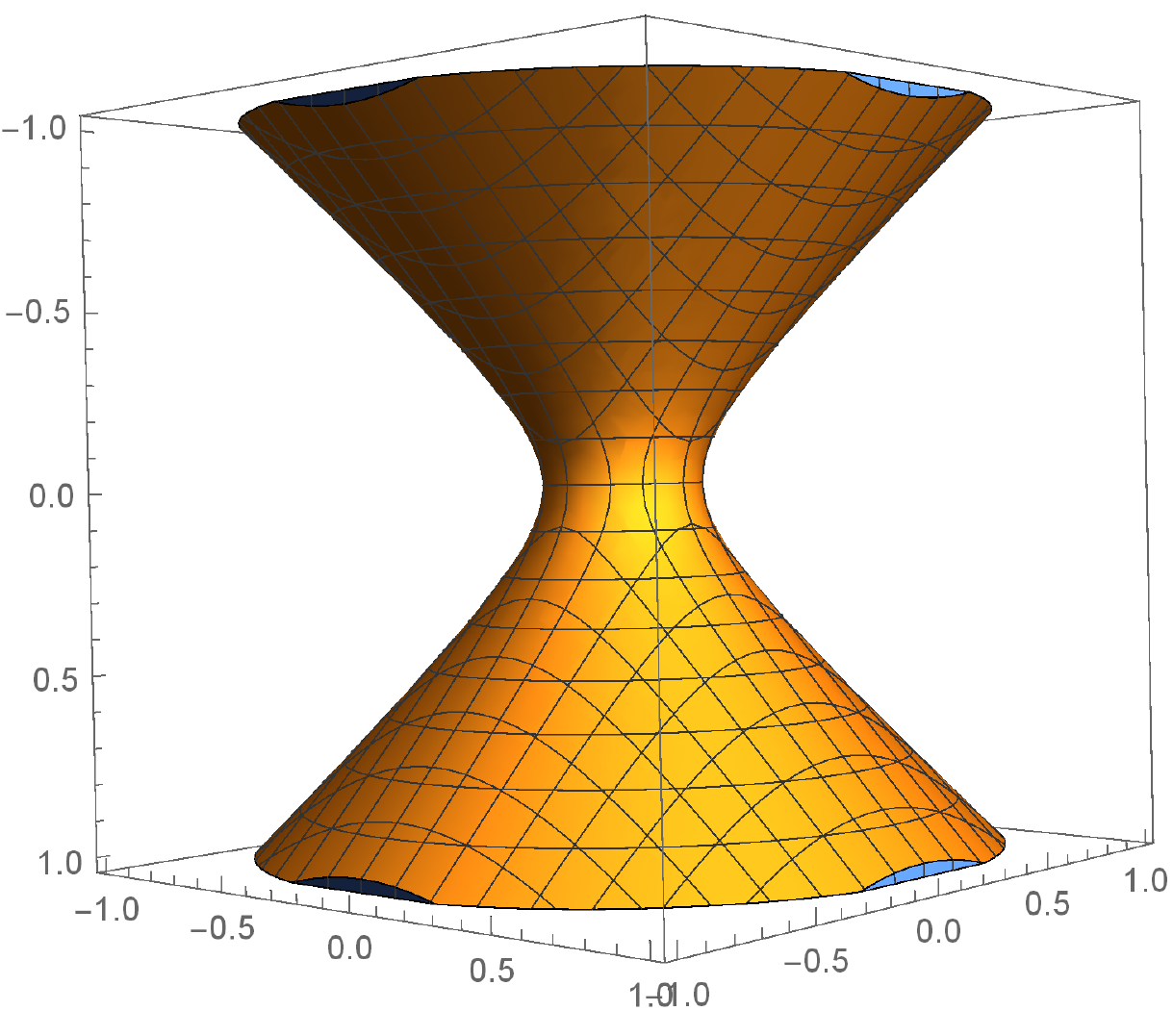}}  
	\hspace{30mm}
  \subfigure[]{\label{energy1}\includegraphics[width=0.20\textwidth]{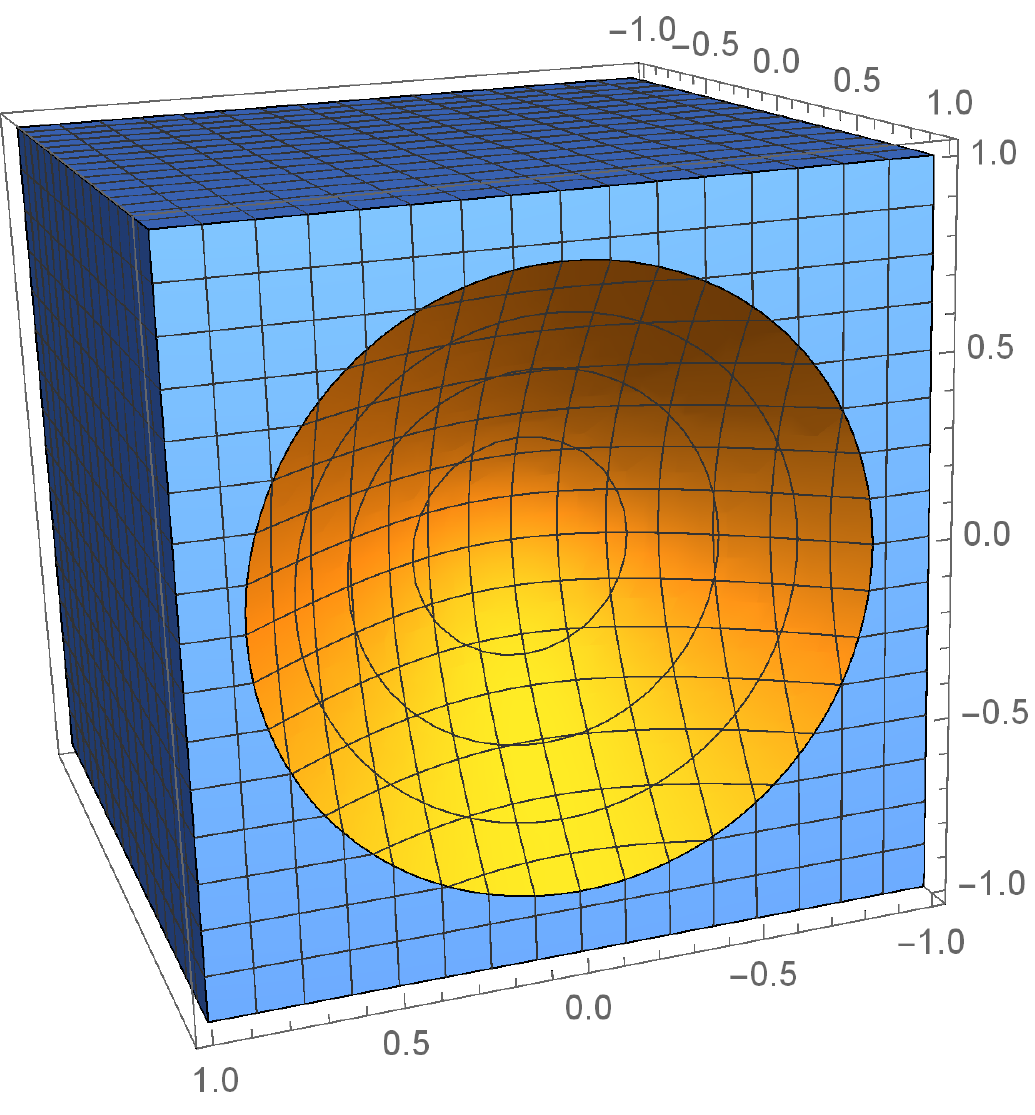}}
 \caption{(a) wormhole in flat space (we have set $\theta = \pi/2$ and used $\phi=\frac{D}{r^2+a^2}$ to obtain a 3-dimensional plot i.e. $\theta = \pi/2$ spatial section of a regular wormhole). Here $a=10$ and $C_0=C_1=C_2=C_3=C_4=0.1$. (b) 3D Plotting of the inequality $ds^2=x_{1}^{2}+x_{2}^{2}+x_{3}^{2}-x_{o}^{2}>0$ for $\theta=0$ (equations (\ref{xi4})). Geodesic retractions of the 5D wormhole space-time $W$ are circles $S_i \subset W$. Here $a=10$ and $C_0=C_1=C_2=C_3=C_4=0.1$}
  \label{Energy}
\end{figure}

\section{Conclusion}
We have investigated the connection between the topological retraction and physical holography. This relation has been clarified through a mathematical restatement to the HP. We have used Euelr-Lagrange equations to perform a geodesic retraction to the 5D Ricci-flat wormhole metric and found that Some types of geodesic retractions are circles. Dealing with the 5D wormhole as hologram, the description of this volume of space is encoded on the lower-dimensional boundary to the region which is a circle. Hence, topological retraction gives an opportunity to explore the possible geometries of the boundary. The simple method used in this article is valid only for Ricci-flat metrics. However, a more general method is required to deal with non Ricci-flat metrics.

\section*{Acknowledgment}
We are so grateful to the reviewer for his many valuable suggestions and comments that significantly
improved the paper.

\end{document}